# Three-dimensional transformable modular kirigami based programmable architected materials

Yanbin Li[1], Qiuting Zhang[1]†, Yaoye Hong[1], Jie Yin[1]*

[1] Department of Mechanical and Aerospace Engineering, North Carolina State University, Raleigh, North Carolina 27695, USA

* Corresponding author. Email: jyin8@ncsu.edu

† Present address: Department of Molecular, Cellular and Developmental Biology, Yale University, New Haven, Connecticut 06520, USA

**Abstract**

Metamaterials achieve unprecedented properties from designed architected structures. However, they are often constructed from a single repeating building block that exhibits monotonic shape changes with single degree of freedom, thereby leading to specific spatial forms and limited reconfigurability. Here we introduce a transformable three-dimensional modular kirigami with multiple degrees of freedom that could reconfigure into versatile distinct daughter building blocks. Consequently, the combinatorial and spatial modular assembly of these building blocks could create a wealth of reconfigurable architected materials with diverse structures and unique properties, including reconfigurable 1D periodic column-like materials with bifurcated states, 2D lattice-like architected materials with phase transition of chirality, as well as 3D quasiperiodic yet frustration-free multilayered architected materials in a long-range order with programmable deformation modes. Our strategy opens an avenue to design a new class of modular reconfigurable metamaterials that are reprogrammable and reusable for potential multifunctionality in architectural, phononic, mechanical, and robotic applications.

**Keywords:** mechanical metamaterials, modular kirigami, reconfiguration, architected materials, programmable deformation## INTRODUCTION

Architected metamaterials are artificial materials constructed through rational design of architectural structures and spatial arrangement of building blocks (1-12) to achieve unprecedented physical properties, such as auxeticity (5, 13), programmability (6, 12, 14, 15), flexibility (3, 7), and cloaking (16, 17). Reconfigurable metamaterials can transform their shapes under deformation and are often assembled by repeating a single type of flexible building block as a module for structural transformation (3). Modular design is often found in both natural and engineering materials and structures ranging from proteins (18) to classical architectures (19) that are composed of multiple repeats of a few basic modular units. Modularity provides a promising strategy to construct versatile reconfigurable architected materials through the designed modules. The advantage of modular architecture resides in its versatility arising from assembling modules that can be connected and arranged in various combinations, as well as reconfigurability for new structures to facilitate a myriad of diverse functional requirements (18).

The ancient paper art of origami (Zhezhi in China) and kirigami (Jianzhi in China) provides unique ways to construct reconfigurable modular architected materials through designing folding and/or cutting patterns of origami and kirigami modules as basic building blocks. Stacking and/or tessellating modular origami (20-23) or kirigami (24, 25) generate transformable periodic architectures through coherent deformation modes of composed reconfigurable modules such as rigid rotation and other deformation mechanisms, which largely expand the design space of architected materials. However, most designs are based on the spatial arrangements of a single type of repeating building block that has limited degrees of freedom (DOF) (20, 21, 24), generating metamaterials with specific structural form and limited properties. The benefits of modular versatility and reconfigurability in architected materials remain largely unexplored, concerning the design of reconfigurable modular building blocks with multiple DOF, as well as the resulting tremendous design space arising from the vast combinatorial assembly and disassembly of their versatile reconfigured modular architectures.

Here, we propose a three-dimensional (3D) transformable modular kirigami strategy to create a new class of reconfigurable, re-assemblable modular architected materials with programmable mechanical response. Far beyond conventional two-dimensional (2D) kirigami through thin sheet cutting or folding with single DOF (25-29), the multi-DOFs 3D modular kirigami is based on closed-loop hinged cubes after cutting (Fig. 1a), allowing transforming in three dimensions into versatile daughter modules in distinct configurations (Fig. 1b). By utilizing the complementary topological features between transformed modules and their vast combinatorial arrangements via stacking and tessellation without bonding connections, we build a library of reconfigurable, re-assemblable periodic one-dimensional (1D) column-like, topologically programmable 2D lattice-like architected materials, as well as 3D quasiperiodic multilayered architected materials with a unique long-range-order. We find that the kirigami-inspired modular assembled 3D architected materials manifest a 3D auxetic behavior with both local deformation independence in each layer and programmable global cooperative deformation. This research paves a new way to design a wealth of both periodic and quasiperiodic architected metamaterials with unprecedented material properties adaptive to diverse applications through combinatorial modular assembly and disassembly.

**RESULTS**

**Transformable 3D Modular Kirigami**

Fig. 1A illustrates the design of a transformable 3D modular kirigami by following two procedures. (i) Starting from a 3D cuboid with dimensions of $l = 2w = 4h$ (*l*: length, *w*: width, *h*: height), we introduce four cutting planes to dissect the cuboid into eight identical cubes with size of *h* connected through eight elastic torsional hinges (non-through cut) that form a closed-loop mechanism, allowing rotation of rigid cubes around hinges system to transform in three dimensions, which reconfigures largely beyond in-plane rotation of rigid plates in conventional 2D kirigami (26-28). (ii) Extruded cut along the *z*-axial direction generates connected thin-walled cubes with uniform orientations defined by the extruded direction, enabling both rigid rotating and soft shearing modes of the thin-walled-cube unit.

When surface contact is allowed during reconfiguration, the 3D modular kirigami exhibits 2 zero-energy modes by assuming free rotational hinges (see Methods), resulting in a multi-DOF structure with its number of DOF $N_{DOFs}$ being $1 \leq N_{DOFs} \leq 2$ accompanied by motion path bifurcation (fig. S1). The kinematic motion path of reconfiguring the modular kirigami can be predicted by a transformation matrix that depicts the cubes' relative positions (30) with (Supplementary Materials)

$$\boldsymbol{n}_j = \boldsymbol{T}_{ji} \cdot \boldsymbol{n}_i \quad (1)$$

where $\boldsymbol{n}_i$ and $\boldsymbol{n}_j$ denote the space vector of any point on cube *i* and cube *j* numbered counterclockwise in Fig. 1A respectively. $\boldsymbol{T}_{ji}$ is their position related transformation matrix determined by the dihedral angles $\gamma_{ij}$ ($1 \leq i, j \leq 8$) of adjacent cubes and their side length (fig. S1), e.g., $\gamma_{12}$ represents the dihedral angle between connected cube 1 and cube 2 as illustrated in Fig. 1A Specially, when setting the dihedral angle $\gamma_{ij} = k\pi/2$ ($k = 0, 1, 2$), Fig. 1B shows that it can reversibly transform into eight different representative configuration states in terms of cube orientation, hinge positions, and transition energy map (fig. S2-fig. S4, Supplementary Materials). The eight configurations are realized through three kinematic transition paths by folding the hinge pair(s) sequentially (Transition-1, $1 \leq N_{\text{DOFs}} \leq 2$, State i – State vi), selectively (Transition-2, $N_{\text{DOFs}} = 1$, State vii), or simultaneously (Transition-3, $1 \leq N_{\text{DOFs}} \leq 2$, State viii) (Movie S1). Interestingly, during transition 3, as the dihedral angle $\gamma_{12}$ increases from 0 to $\pi/2$, both nominal normal strain $\varepsilon_{yy}$ and $\varepsilon_{zz}$ along the $y$ and $z$ axis increase monotonically while $\varepsilon_{xx}$ exhibits a peak (inset of Fig. 1C). Consequently, its Poisson's ratios $v_{xy} = -\varepsilon_{xx}/\varepsilon_{yy}$ and $v_{xz} = -\varepsilon_{xx}/\varepsilon_{zz}$ transit from negative to positive values with a turning jump point at the peak nominal strain $\varepsilon_{xx}$ (Fig. 1C) caused by the unique structural form (fig. S2, Supplementary Materials).

Moreover, if the corner creases of the cubes are flexible, we note the basic unit can exhibits different soft modes under different transition states. Specially, selecting the state-ii and the intermediate state (Fig. 1D, right-top) between state-ii and state-iii as representatives, we demonstrate their soft modes by shearing cubes with same orientations (Fig. 1D, bottom), and by the following utilize them to achieve unique material properties for our designed multi-dimensional architected materials.

Interestingly, we note that such a 3D kirigami modular design could further evolve into 31,845 derived basic modules as potential building blocks through combinatorial designs of both relocated hinge connections (for example in Fig. 1E-i by fixing the side four hinges and relocating the remaining 4 hinges) and different thin-walled cube orientations (for example in Fig. 1E-ii by tuning the middle four cube orientations highlighted with green color), see details in Methods. The combinations of such eight building blocks in terms of cube orientations and spatial hinge arrangements will provide enormous design space for constructing reconfigurable, re-assemblable, reprogrammable, multi-dimensional architected metamaterials through modular assembly and disassembly.

**Reconfigurable 1D and 2D Architected Materials**

Through vertical stacking and in-plane tessellation of derivative transformable building blocks, we can generate 1D column-like (Fig. 2) and 2D lattice-like (Fig. 3) reconfigurable periodic architected materials, respectively. We fabricate the 3D modular kirigami prototypes on a centimeter scale made from cardboards by exploiting laser-cutting and stepwise bonding technique (see Methods, fig. S6).

During modular assembly, two geometrical compatibility conditions need to be satisfied to ensure coherent deformation and structural reconfiguration (11). First, at shared boundary surfaces, pairs of neighboring blocks should display complementary shapes, such that adjacent blocks can be adapted for tight fit without protrusions regardless of further deformations (e.g., Fig. 2A-i, Fig. 2B-i, and Fig. 3A-i). Second, regarding the in-plane alignment and tessellation of blocks, all the blocks should fit with matched complementary topologies to ensure cooperative deformation among all units (e.g., Fig. 3A-i).

Guided by the above compatible conditions, Fig. 2A-iii and Fig. 2B-iii present two examples of assembled 1D column-like compact materials through complementary shapes between two derivative transformable building blocks without bonding boundary faces of adjacent units: one is constructed by directly stacking the single transformed module of State viii (Fig. 2A-ii); the other is by orthogonally stacking the same transformed module of State vii (Fig. 2B-ii). The former is reconfigurable and deforms collectively the same as its constituent single module (Fig.

2A-i), i.e., closing the void in the center under lateral compression through rigid mode or shearing the closed square column through soft mode (Fig. 2A-iii). By contrast, the latter shown in Fig. 2B-iii is rigid and non-deformable due to the heterogeneous cube orientations and surface contacts; however, it can reconfigure in soft shearing mode by setting all cubes with uniform orientations (fig. S7).

When line bonding connections between modules are allowed, more varieties of reconfigurable assembled 1D column-like materials could be generated through kinematic bifurcation. Fig. 2C-i shows that stacking the same modules as Fig. 2A-i but bonding the four pairs of edges (highlighted in yellow dashed lines) with neighboring modules generates a 1D column-like material with gaps between layers, where the opening angle $\theta = \pi/2$ denoted in Fig. 2c-i. Similarly, the assembled structure can be further opened or closed under lateral compression with $0 < \theta < \pi$ through the rigid rotation mode, exhibiting a single-DOF deformation mode (Fig. 2C-ii). However, once it is closed, i.e., $\theta = 0$ or $\pi$, it will reconfigure and bifurcate into a multi-DOF compact structure with a number of combinatorial deformation mobility $N = 2^{n+1} - 1$ (Fig. 2D), where $n$ is the stacking layer number. For a 5 layered structure, it could have a total of 63 different reconfigured states with 2 of them shown in Fig. 2C-iii.

Rather than stacking, tessellation of State-viii module with in-plane complementary shapes along two orthogonal directions (Fig. 3A-i) satisfying both aforementioned geometrical compatibility conditions forms a periodic 2D lattice-like material, which takes similar structural form (Fig. 3A-ii) to the conventional kirigami sheet with square units (26-28). It shows a constant negative Poisson's ratio of −1 by allowing only the rotation mode of rigid assembled cubes (highlighted in green color) to close the voids under compression (Fig. 3A-iii). Furthermore, such voids in the 2D materials matrix provide sites to fill structurally compatible modules of State ii or State v as inclusions (Fig. 3B-i, ii), thus generating more varieties of assembled 2D architected materials with tunable materials response by selectively filling or removing the porous matrix with inclusions (Fig. 3B-iii, iv). As one example, in Fig. 3C-i, we set the assembled matrix with uniform cube orientations along the thickness direction. Under compression, it shows a chiral deformation centering around the assembled cube (Fig. 3C-ii). We demonstrate that depending on the cube orientation of inserted modules, the 2D lattice-like material could exhibit topologically distinct structural phase transitions under compression. When the inclusions are placed with their cube orientations orthogonally to the matrix (Fig. 3C-iii), such inclusions become rigid and are much stiffer than the matrix, thus, upon compression, the hybrid material undergoes soft shearing deformation in the matrix accompanied by the rigid rotation of the inclusions. Correspondingly, the chirality center shifts from the original matrix (bottom of Fig. 3C-i) to the inclusions (bottom of Fig. 3C-ii). By contrast, when the inclusions are placed with the same cube orientations as the matrix (top of Fig. 3C-iii), the original deformation chirality is eliminated (bottom of Fig. 3C-iii) and the non-chiral structure shows a positive Poisson's ratio, where the soft inclusions possess the same stiffness as the matrix and the whole structure deforms under soft shearing modes in both the matrix and inclusions.

**Programmable Quasiperiodic 3D Architected Materials**

Design of 3D architected materials can be considered as a spatial filling tessellation problem, where the composed periodic/nonperiodic structures can have voids (weak tiling with gaps) or no vacancies (strict tiling with compactness) between building blocks (31). Periodic metamaterials favor cooperative structural deformation from their geometrically compatible periodic units (3, 10, 21), while in nonperiodic metamaterials, intrinsic incompatible geometric frustrations after deformation can be avoided through appropriate design of building blocks (11). Consequently, to design quasiperiodic 3D architected materials composed of both periodic and nonperiodic units, it needs to address both challenges of the geometric frustration between non-periodic units and deformation compatibility between distinct non-periodic and periodic units.

Inspired by the mortise-and-tenon connection method (Sunmao in Chinese) in ancient Chinese architecture (19), we utilize the conformable fit between different combinations of 3D modular kirigami to create a new class of quasiperiodic yet frustration-free 3D architected metamaterial that possesses a long-range order (32) without bonding connections (Fig. 4A). The 3D metamaterial is formed by directly connecting layers of 2D metamaterials along the $z$-direction with tightly fitted modular kirigami square columns without any bonding, see schematic model and prototype in Fig. 4B. The arrays of square voids in the 2D layer play the role as mortises to join the square columns in different patterns (Movie S2), which are composed of two line-hinged basic modules (State i) by connecting the two pairs of side edges of top and bottom cubes (middle of Fig. 4C). The column can be divided into three segments in terms of connection, where the four cubes on the top and bottom segment highlighted in blue color represent the tenons inserted into the voids of 2D layers, and the eight cubes in the middle segment highlighted in green color represent the column connector with internal structures (middle of Fig. 4c).

The potential deformation modes of the proposed architected materials in a long-range order are determined by both the mortise-and-tenon connectivity and the internal structures of column connectors. Considering the heterogeneously voxelated cubes and conditional hinge positions in the square column, the combinatorial number of evolved structural forms in the column could be over $31,845^2 \approx 10^9$. However, we can readily classify them into $2^3 = 8$ deformation motifs in terms of combinational deformability in the three-segmented column (top tenon, middle column connector and bottom tenon), i.e., (+, +, +), (+, +, −), (+, −, +), (−, +, +), (+, −, −), (−, +, −), (−, −, +), and (−, −, −) ('+' denotes deformable and '−' denotes rigid and non-deformable), see representative examples with stretchable and compressible features in fig. S8 and Movie S3. Fig. 4c shows two representative examples of columns with motifs of (+, +, +) (right) and (−, +, −) (left). We note that the square columns with identical structural forms can deform distinctly by manipulating hinge positions. For both columns, the column connector in the middle can deform in multiple modes through shearing of internal structures under compression. Both top and bottom tenons are stiff in the (−, +, −) column but are deformable through soft mode in the (+, +, +) column. Notably, for (+, +, +) column, we find that both top and bottom tenons can be independently twisted without deforming the middle connector.

To explore the structural response of long-range-order quasiperiodic architecture, we use the polarized spins ($n_{ix}$, $n_{iy}$, $n_{iz}$) to characterize local directional shearing in cubes along respective $x$, $y$, and $z$-axis, and map this ordered quasi-periodic materials onto a directional spin frame (Fig. 4d) that satisfies the so-called "ice-rule" (33) (see Methods). The spin takes values of −1, 0, or +1, where +1 and −1 represent deformable cubes with polarizations corresponding to the positive and negative axial direction, respectively, while 0 implies nondeformable cubes. Thus, we can use the simplified "ice-rule" to describe the programmability of our long-range ordered architectures. For example, the different deformation modes of the $i^{th}$ layer consisting of both 2D materials and the (+, +, +) columns shown in Fig. 3d can be characterized by the spin configurations with different spin values. Basically, we can uniquely correlate each admissible spin configuration to a specific local deformation mode for any allowable structural form of voxelated column. Notably, for the $i^{th}$ layer with $(n_{ix}, n_{iy}, n_{iz}) = (\pm 1, \pm 1, \pm 1)$, both 2D materials and the column can deform independently (Fig. 4e).

Consequently, the arrangement of spin configurations controls the number of deformation modes of the assembled architected materials. To this end, we assemble (+,+,+) columns into a three-layered 3D architectures with spin configurations for all layers as $(n_{ix}, n_{iy}, n_{iz}) = (\pm 1, \pm 1, \pm 1)$ and fabricate the physical samples by assembling three 6×6 units-based 2D material layers connected with two layers of 2×3 inserted columns (Fig. 4f and Movie S4 and Movie S5). With the deformation independence between 2D layers and column connectors, the designed 3D architected materials possess multiple DOFs to achieve unique materials response. We demonstrate that each 2D material layer can deform independently and locally under compression through cube rotation to close the voids without deforming other columns and 2D materials despite the connections (Mode 1, inset of Fig. 4g). In addition, it can be sequentially deformed by laterally compressing the top layer first, followed by column compression through cube shearing (Fig. 4g). Furthermore, we can achieve globally cooperative deformation modes in the whole structure characterized by macroscopic nominal strains $\varepsilon_{xx}$, $\varepsilon_{yy}$, and $\varepsilon_{zz}$ along $x$, $y$, and $z$-axis, respectively, including $\varepsilon_{xx} = \varepsilon_{yy} = 0$, $\varepsilon_{zz} \neq 0$ (Mode 2, Fig. 4h) under vertical compression through local shearing of columns, $\varepsilon_{xx} = \varepsilon_{yy} \neq 0$, $\varepsilon_{zz} = 0$ (Mode 3, Fig. 4i) under lateral compression through rigid rotation in 2D materials, and $\varepsilon_{xx} = \varepsilon_{yy} \neq 0$, $\varepsilon_{zz} \neq 0$ (Mode 4, Fig. 4j) under both vertical and lateral compression through both cooperative column shearing and cube rotation in 2D materials (see also Movie S6b). Consequently, it results in a 3D auxetic material with a constant negative Poisson's ratio of $v_{xy} = v_{yx} = -1$ and zero Poisson's ratios of $v_{xz} = v_{zx} = v_{yz} = v_{zy} = 0$ in all other planes with $v_{ij} = -\varepsilon_{ii} / \varepsilon_{jj}$.

Based on the deformation independence between 2D layers and columns, for a long–range ordered quasi-periodic metamaterial with a total number of $N$ layers of 2D materials and ($N$–1) layers of columns, we could obtain $2^{2N-1} -1$ ($N \geq 1$) (Supplementary Information) disparate combinatory deformation modes, providing a potential enormous design space to tune structural forms and programmable material responses. For example, when $N = 10$, it could generate over $5.2 \times 10^5$ different deformed configurations (Fig. 4k), making this material ultra-programmable.

We note that the topological properties of our designed 3D architecture, such as structural anisotropy and deformability, can also be tuned through divergent combinatorial deformation modes of different column types made of distinct cube voxels (fig. S9, Supplementary Materials, Movie S6a). Notably, arising from deformation independence for this ultra-programmable architecture, we can achieve surprisingly quasiperiodic (column layers with distinct structural forms) to periodic transition, i.e., the 3D structure transforms from original quasi-periodic with a long-range order to a periodic monoclinic crystal-like lattice identified with certain lattice invariances (fig. S10) after a complete compression ($\varepsilon \approx 100\%$) in all column layers, a new phenomenon in metamaterials not demonstrated before.

**Mechanical Properties of 3D Architected Materials**

Finally, we show that the material response of our proposed 3D architected materials can be manipulated by the multiple deformation modes in the columns. Fig. 5a shows the cardboard prototype of 3D architected materials assembled through three layers of lattices ($2 \times 2$ units, green color) bridged by two layers of four square columns.. Its mechanical response is investigated under uni-axial compression (fig. S11-S12, Methods). The column type chosen here has the same hinge arrangements as Fig. 4c-i but with homogeneous cube orientations in the middle segment, resulting in a spin configuration of $(n_{ix}, n_{iy}, n_{iz}) = (0, \pm1, 0)$. Interestingly, we find that the cardboard-based columns exhibit excellent elastic recovery after several hundred cycles of loading/unloading tests (fig. S11e and fig. S12a) for their potential reusability.

We note that the architected material demonstrates multi-mode and nonlinear deformation characterized by the distinct folded structures induced by independent local shearing in the columns, which lead to four compression modes from Mode 1 to Mode 4 and one interesting shearing mode (Fig. 5b). For each mode, we conducted the compression tests of the prototype for over 350 times (see Supplementary Materials) and calculated stiffness $k$ by the slope of force-compression curves (Fig. 5c-5d). Overall, we can differentiate the exhibited averaged *J*-shaped force-compression curves of cardboard samples into two stages (Fig. 5c): an initial linear-elastic stage over a long range attributed to dominated local shearing in the voxelated cubes through hinge-rotation and then transit to a steep strain hardening stage due to cube wall crumpling (see Movie S8-9). We observe that the stiffness $k$ for the four compression modes is almost the same but much higher than the shearing mode (Fig. 5d and Movie S10). We believe that such stiffness decrease in shearing mode is due to the deformation-coupling reduction between the hinge rotations and released bending of cube walls by the misaligned deformation of middle 2D material layer. By setting the top column layer with lower torsional hinge stiffness through engraving in the carboard sample (fig. S13), we demonstrate a sequential deformation mode (bottom right of Fig. 5b) with exhibited four distinct stages in the force-displacement curves (Fig. 5e). The top layer of columns is compressed first through cube shearing due to its lower torsional stiffness in the hinges, i.e., a smaller slope in Stage I, followed by the bottom column layer with higher hinge torsional stiffness, i.e., an increased slope in Stage II. Compression of both column layers leads to a further increase in the slope in Stage III followed by a steep rise in Stage IV due to densification. Such a sequential deformation mode could provide more tuning space for programmable multi-step material response especially for multilayer architecture materials.

Furthermore, we also fabricate the same 3D architected metamaterials through multi-material 3D printing (Fig. 5f, Methods), where the cube walls in white color are printed with rigid materials and quarter-arc-shaped soft line hinges in black color are printed with soft rubber-like materials. It serves two purposes: one is to demonstrate the applicability and capability of efficiently fabricating the modular assembly of our proposed architected materials without sacrificing reconfigurability; the other is to eliminate the wall bending in the cardboard prototypes by setting a relatively thicker wall in the printed prototype. Similar to the assembled cardboard-based metamaterials, the 3D printed architected materials demonstrate programmable and independent deformation modes in the printed column, as well as realize divergent combinatorial structural forms by independently or dependently deforming the column or 2D lattice layer (Fig. 5g, fig. S13). We note that the 3D printed sample exhibits highly nonlinear distinct *S*-shaped curves with three stages for both compression and shearing deformation modes (Fig. 5h), where the shearing mode leads to a much lower stiffness. Interestingly, the steep strain hardening stage under large compression shows similar stiffness with the initial curve. We attribute the exceptional curve of the printed 3D architecture to the dominated bending in the soft hinges that undergo centroid or off-centroid curve-beam bending during different compression stages (Movie S11-12).

**DISCUSSION**

We introduced a 3D modular kirigami approach to design versatile multi-dimensional architected metamaterials constructed from combinatorial assembly of the same multi-DOF module but in different transformed configurations. The closed-loop connection of 8-hinged cube module provides a multi-DOF mechanism to transform into versatile daughter modules. Specially, the complimentary transformed configurations between different daughter modules enable assembly, disassembly, and re-assembly into varieties of reusable or new forms of architected materials (Movie S13) for materials sustainability and next level of reconfigurability, which could not be achieved in most state-of-art *non-disassemblable* architected materials (1-4). We envision that such a closed-loop connection mechanism could be applied to different number of connections and other connected prismatic, tetrahedral, and polyhedron shaped units or their combination for constructing more varieties of 3D architected materials with or without bonding connections. In this work, we only explored a small portion of representative structures showing unique materials response. We believe that the enormous combinations of encoded voxelated cubes with polarized shearing directions and their conformable connections between different modules in each assembled component will largely expand the design space of reconfigurable modular kirigami-inspired architected materials. The study of these novel materials will provide insight toward the implementation of deformation-controlled, reprogrammable, reusable, and multifunctional materials for minimizing material waste and a range of architectural, phononic, electrical, mechanical, and robotic applications.

**METHODS**

**Fabrication of single cardboard cube with soft shearing mode.** The fabrication process of a single cardboard cube is shown in fig. S6 by following three steps. First, based on the prescribed cutting and engraving patterns for generating respective line cuts and creases in the cardboard, we cut the patterned rectangular strip out with a laser cutter. Second, we covered the non-engraved side of the strip with plastic tape to enhance its reusability and good material elasticity in both the line hinges and thin walls. Third, we sealed the strips at the overlapping end with double-sided tapes and folded the creases by 90º to generate the cardboard cube, see more details in the Supplementary Materials.

**Combinatorial design of evolved 3D modular kirigami.** We treat the sublevel eight thin-walled cubes in the basic module as anisotropic voxels with extruded orientations. The basic module of 8 cubes with a closed-loop eight-hinged connection could evolve into 31,845 basic modules through combinational design of the orientations of 8 cubes (each cube possessing three polarization directions along *x*, *y*, and *z*-axis for shearing leads to 6,369 combinations, see fig. S5b and Supporting Information), as well as relocated hinge connections (each module has at least 5 connection ways of different hinge positions, see fig. S5).

**Zero energy mode for 3D modular kirigami.** When the elastic line hinges between truncated sublevel cubes are completely flexible, by using the Maxwell counting argument, we can determine the number of zero modes (free rotation of individual cubes without energy stored in the system). We begin with six degrees of freedom per cube (three directional translations, three rotational spins) and subtract three constraints from each connection, meanwhile it has six global degrees of freedom (global translation and rotation). For a system with *n* cube and *m* line-hinge connections, the internal number of degrees can be expressed by means of Grübler's Formula (34)

$$N = 6(n - m - 1) + \sum_{i=1}^{m} f_i \qquad (2)$$

wherein $n = m = 8$ for the designed 3D modular kirigami, $f_i$ is the number of degrees of freedom associated with the $m^{th}$ elastic line hinge which is equal to 1. Thus, for the 3D modular kirigami, it yields 2 internal degrees of freedom, and some of them with different configurations are shown in Fig. 1b.

**Reduced Ice-rule model.** We assume that all the individual thin-walled cubes are composed of four rigid plates that can perform rotational deformation through elastic flexible hinges. The spins frame shown in Fig. 4d is a reduced simplified model since we only consider the z-directional spin and assume plane connection between spins. Each column is represented by a vertex and the 2D layers are denoted by parallel planes; then, we connect these vertices and planes with line bonds along long-range order direction to represent overlapping cubes between 2D material layer and columns. However, due to the special structural connectivity of our quasi-periodic 3D architecture with compatible conditions restricted by equations (2) and (3), only spin $n_{iz}$ can be placed along bonds while spins $n_{ix}$ and $n_{iy}$ of the middle column part must be fixed at corresponding vertices (Fig. 4d).

**Fabrication of 3D printed architected materials.** The commercial multiple material 3D printer Stratasys Objet 260 Connex3 is utilized to fabricate the 3D printed architected materials. To realize the hinge rotation mechanism, we simplify the line hinges with printed bendable soft hinges represented by the black curve part (fig. S14a) with materials of Agilus30Black–FLX 95550–DM. The thin walls of voxelated cube are printed with stiff materials (RGD 5131–DM). The thickness and length of hinges is 1 mm and 10 mm with 1mm inner radius and 2mm outer radius. The thickness and length of the rigid walls are 1 mm and 10 mm, respectively (see the schematics in fig. 14a and more details in Supplementary Materials).

**Mechanical Test**. We conducted uniaxial compression tests for columns with different voxelated cube patterns at a rate of 0.01 mms$^{-1}$ (Instron model 5944) to characterize the relation between the compressive force $F$ (N) and the compression displacement $u$ (mm) (see more details in Supplementary Materials).

## SUPPLEMENTARY DATA

Supplementary data are available at NSR online.

## FUNDING

This work was supported by National Science Foundation under award number CMMI CAREER-2005374.

## AUTHOR CONTRIBUTIONS

Y.L. and J.Y. proposed and designed the research. Y.L. designed and fabricated the prototypes. Y.L. performed theoretical analysis of the system. Y.L., Q. Z., and Y.H. performed the experiments. Q.Z. performed the numerical simulation. Y.L. and J.Y. wrote the manuscript.

## COMPETING INTERESTS

The authors declare that they have no competing interests.

**Figures**

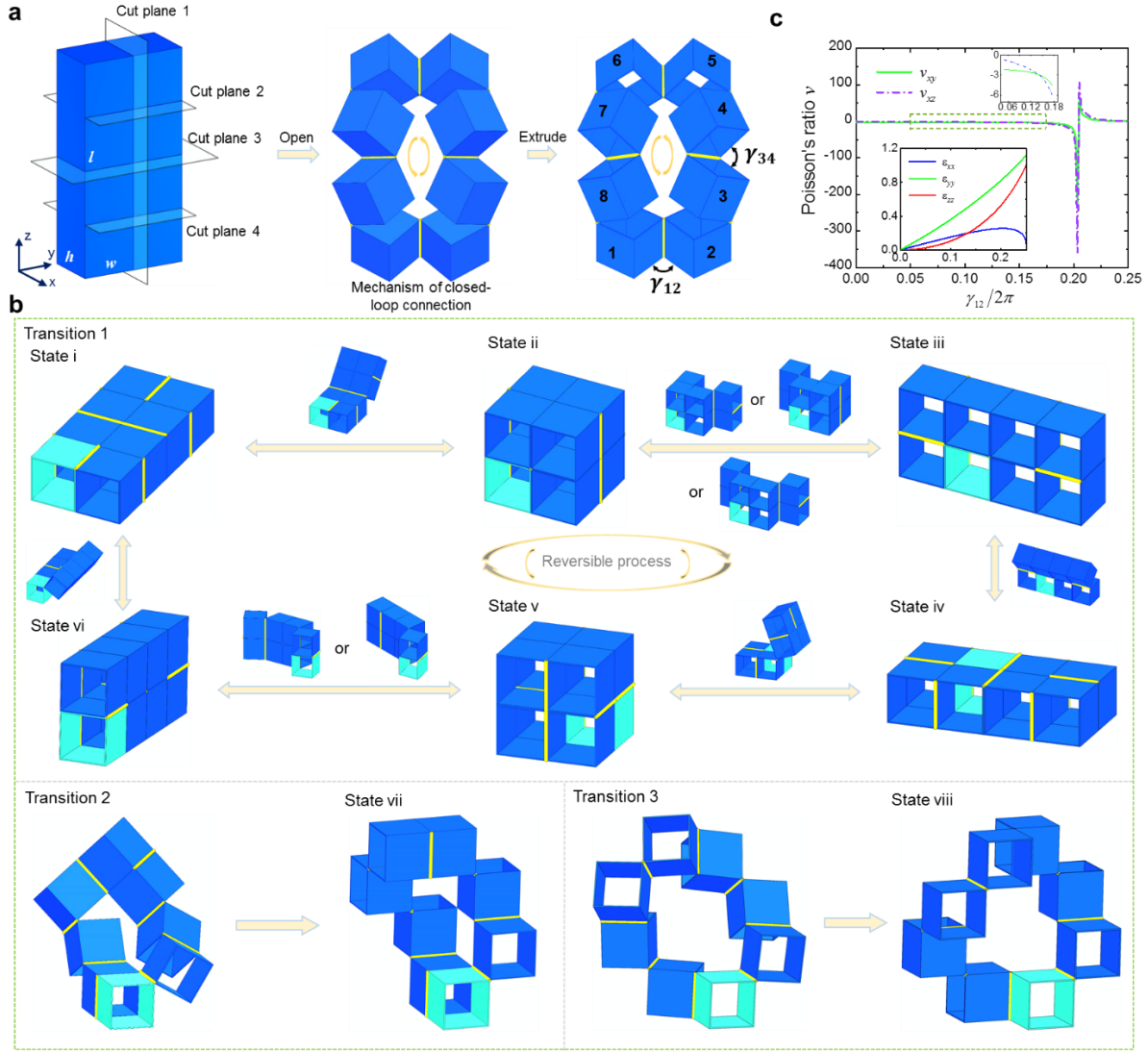

**Figure 1: 3D transformable modular kirigami.** (**a-b**) Schematic construction of a 3D modular kirigami (**a**) and its eight representative transformed configuration states through three different reversible kinematic transition paths (**b**). In (**a**), a solid cuboid is first dissected by four cutting planes into eight identical cubes connected through elastic torsional hinges (highlighted in yellow lines) to allow reconfigure in 3D by rigid rotation in a closed-loop mechanism, followed by extruding to generate connected thin-walled cubes, allowing deforming extruded cubes through thin wall shearing. $\gamma_{ij}$ is the dihedral angle between two adjacent cubes $i$ and $j$. In (**b**), $\gamma_{ij}$ is set to be $\gamma_{ij} = k\pi/2$ ($k = 1$ or 2) after reconfiguration. (**c**) profile of Poisson's ratio $v_{xy}$ and $v_{xz}$

versus opening angle $0 \leq \gamma_{12}/2 \leq \pi/2$ during Transition 3 in (**b**). Top inset: zoomed view of Poisson's ratio vs. $\gamma_{12}/2$ profile. Bottom inset: profile of nominal strains along three axial directions versus $\gamma_{12}/2$.

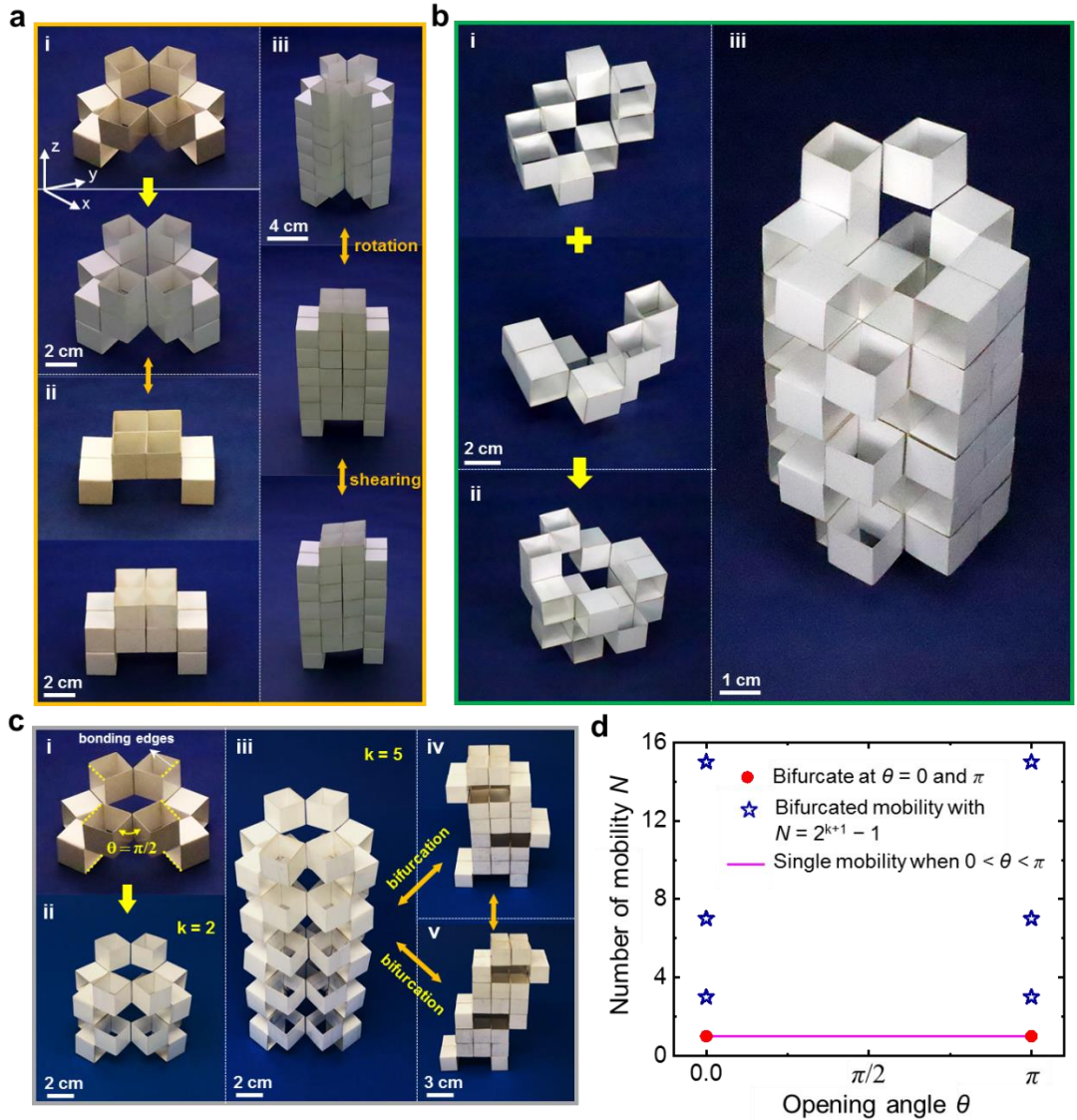

**Figure 2: Periodic 1D column-like architected materials through modular assembly.** (**a-c**) Prototypes of cardboard-based 1D column-like metamaterials by vertically stacking the same building blocks without bonding (**a**-**b**) and with bonded edges (**c**). The repetitive building block is composed of single module of transformed State viii in (**a**) and (**c**), two complimentary modules from the same transformed State vii in (**b**). The created 1D metamaterial is reconfigurable with single DOF through rigid rotation or soft shearing mode to close or open the pores (**a**), while it is nondeformable in (**b**) with both deformation modes locked. In (**c**), when it is closed with the opening angle θ = 0 or π, it bifurcates into a multi-DOF structure with its number

of mobility shown in (**d**). Two selected transformed configurations are shown in c-iv. (**d**) The number of mobility in the 1D assembled metamaterials of (**c**) as a function of $\theta$ and stacking layer number $k$. Intermediate states with ($0 < \theta < \pi$) have single mobility.

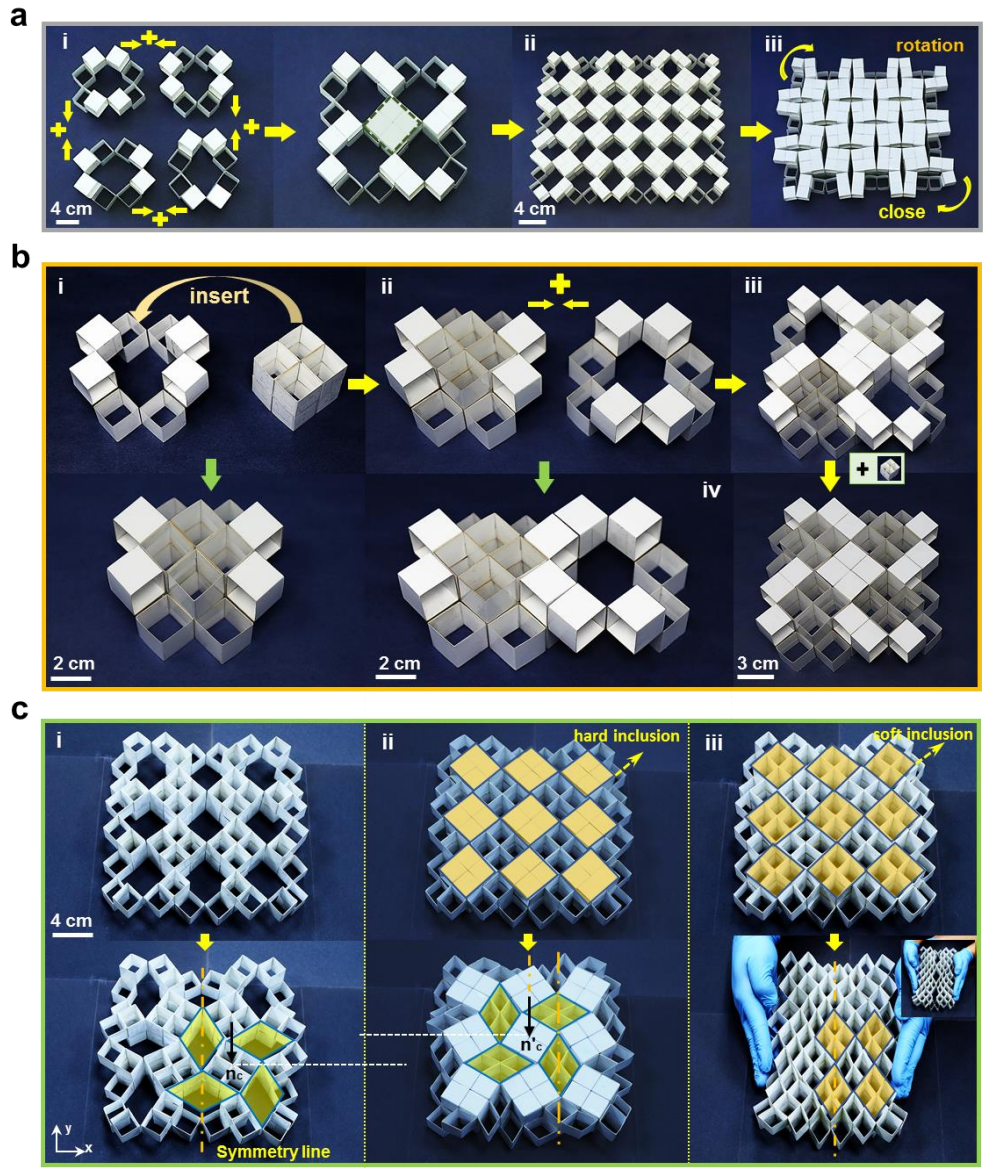

**Figure 3: Periodic 2D lattice-like architected materials through conformable modular assembly.** (**a**) 2D lattice-like metamaterials by in-plane tessellation of transformed modules with complimentary topological configurations. The repetitive building block is composed of four complimentary modules from the same transformed State viii. It forms reconfigurable 2D metamaterials through rigid rotation mode to close the pores. (**b**) Programmable 2D metamaterials by selectively filling the voids of tessellated modules of State viii with tightly fitted cubic modules of State ii or v as removable inclusions. (**c**) Tunable deformation chirality in 2D metamaterials through the same filled inclusions but different cube orientation under uniaxial compression. (i) 2D metamaterial as matrix composed of assembled State viii modules with uniform cube orientation exhibits deformation chirality centering at matrix ($\boldsymbol{n}_c$). (ii) After filling the matrix with hard inclusions that possess different cube orientation, the deformation chirality center shifts from the matrix to the rigid inclusion ($\boldsymbol{n'}_c$). (iii) It transits to a non-chiral structure after filling the matrix with soft inclusions that possess the same cube orientation.

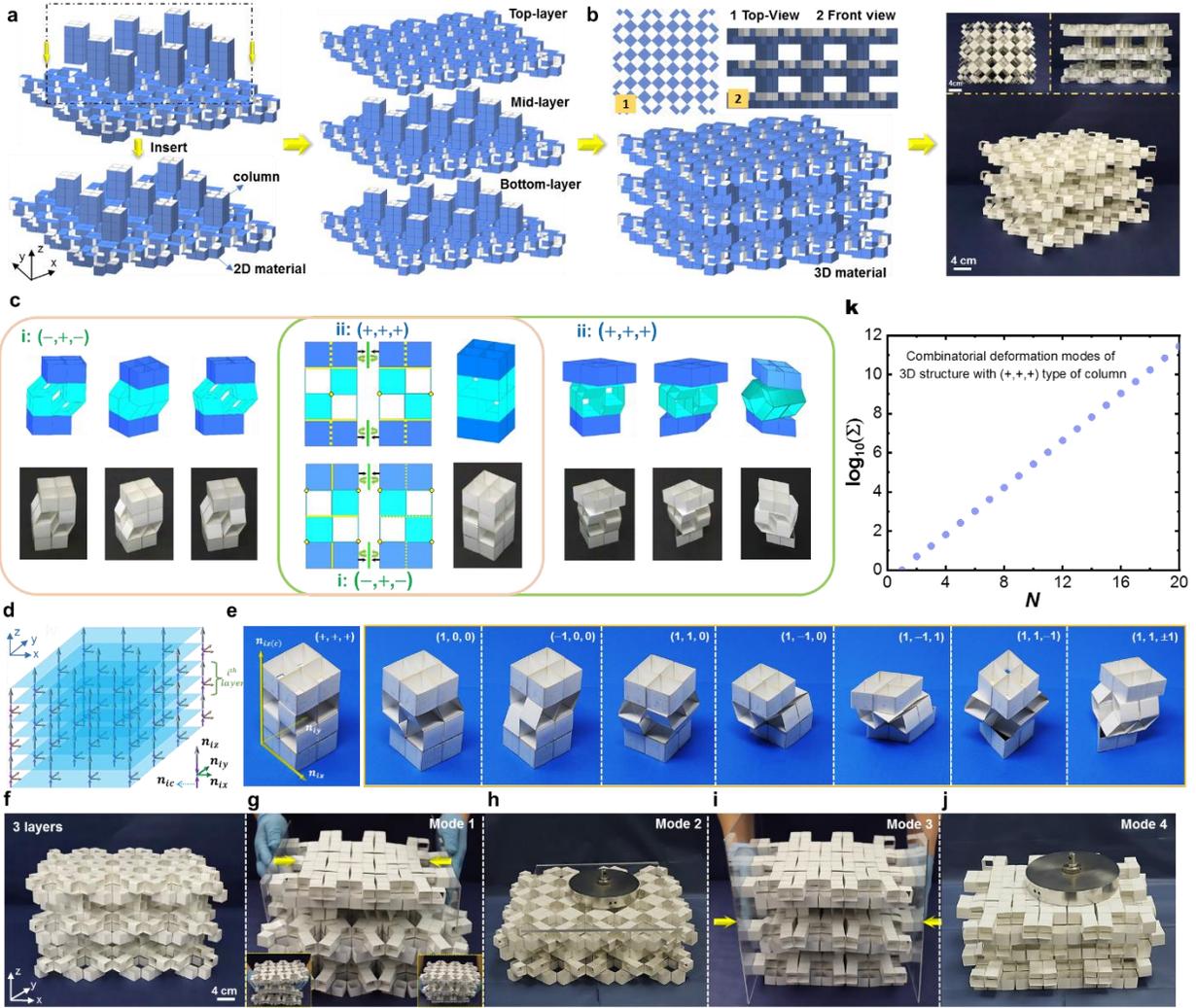

**Figure 4: 3D quasiperiodic architected metamaterial.** (**a**) Schematic construction of multilayered architected metamaterials via Chinese mortise-and-tenon connection technique by filling the voids (mortises) of 2D assembled material layers (same as Fig. 3a(ii)) with inserted arrays of tightly fitted square columns (two ends as tenons). (**b**) Schematic model (left) and corresponding prototype of cardboard-based architected material (right) in different views through modular assembly. (**c**) Two representative designs of square columns composed of two line-hinged basic kirigami modules (State i) by manipulating hinge positions (middle). The two columns have identical structural forms but different deformation modes in the top, middle, and bottom segment. Column type 1 (−, +, −) can only deform its middle segment through shearing

of the voxelated cubes (left). In column type 2 (+, +, +), all the three segments can deform independently (right). (**d**) Mapping of the assembled 3D architectures to the simplified spin frame, where ($n_{ix}$, $n_{iy}$, $n_{iz}$) are rotation spins along the *x*, *y*, *z*-axis direction, respectively. $n_{ic} = n_{iz}$ is the rotation spin in the 2D material layer. (**e**) Combinatorial deformation modes for column type 2 in (**c**) characterized by different values of rotation spin ($n_{ix}$, $n_{iy}$, $n_{iz}$). (**f**) Prototype of cardboard-based three-layered architected metamaterial composed of three layers of 6 × 6 2D materials and two layers of 2 × 3 square columns (the same column arrangement as (**a**) with spin configuration of ($n_{ix}$, $n_{iy}$, $n_{iz}$) = (±1, ±1, ±1). The composed square column (+, +, +) is the same as (**e**). (**g**) Demonstration of its local independent deformation mode between each 2D materials layer and square columns layer (Mode 1) in (**f**). (**h-j**) Demonstration of its global cooperative deformation modes under vertical compression by deforming columns only (Mode 2, **h**), under lateral compression by deforming 2D materials layers only (Mode 3, **i**), under combined vertical and lateral compression by deforming all layers (Mode 4, **j**). (**k**) Total number (Σ) of combinatorial deformation modes for 3D architected materials with the number of *N* layers of 2D material parts bridged with column type (+, +, +).

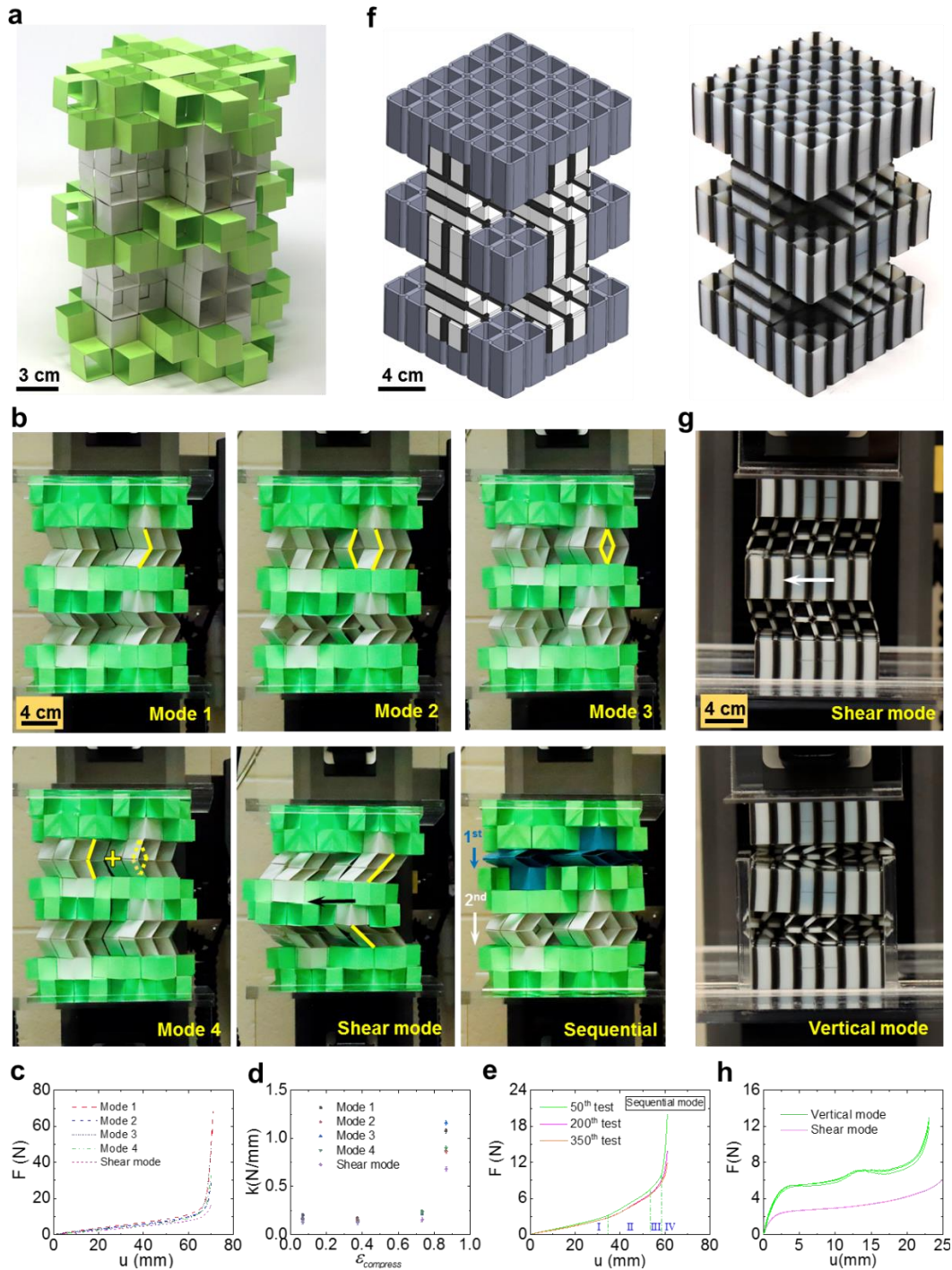

**Figure 5: Mechanical properties of 3D architected materials.** (**a**) Prototype of cardboard-based three-layered architected materials composed of 2 × 2 2D materials and 2 × 2 columns with deformation motif (−, +, −) through modular assembly (see Supplementary Video 13 for disassembly process). (**b**) Six different deformation modes in cardboard-based 3D architected

materials under compression in terms of combinatorial directional shearing in columns: four vertical compression modes (Mode 1, 2, 3, and 4), one shear mode, and one sequential deformation mode. (**c-e**) Corresponding mechanical properties characterization of the cardboard-based prototype under different deformation modes. (**c**) Experimental compression force F vs. displacement u curves under compression Modes 1 – 4 and shear mode by averaging 350 compression tests. (**d**) Corresponding measured structural stiffness under the five deformation modes at four compression strains of 6.7%, 37.3%, 73.3%, and 86.7%. (**e**) Experimental force-displacement curves under sequential deformation mode at $50^{th}$, $200^{th}$, and $350^{th}$ test, showing four-stage deformation and nearly repetitive elastic deformation in the cardboard-based protype. (**f**) Corresponding schematic model (left) and prototype through direct 3D multi-materials printing (right). Soft hinges are highlighted in black color and hard cube faces are highlighted in white color. Both prototypes possess spin configuration of $(n_{ix}, n_{iy}, n_{iz}) = (0, \pm 1, 0)$ (see fig. S8b). (**g**) Corresponding shear mode and vertical compression mode in the 3D printed prototype. (**h**) Experimental force-displacement curves of 3D-printed prototype under shear and compression mode.